\documentclass[twocolumn]{aastex631}

\usepackage{xcolor}
\usepackage{xurl}

\received{March 3, 2024}
\accepted{April 16, 2024}

\submitjournal{PASP}

\begin{document}

\title{The Robotic MAAO 0.7\,m Telescope System: Performance and Standard Photometric System}

\correspondingauthor{Gu~Lim and Dohyeong~Kim}
\email{lim9gu@gmail.com; dh.dr2kim@gmail.com}

\author[0000-0002-5760-8186]{Gu~Lim}
\affiliation{Institute for Future Earth (IFE), Pusan National University, Busan 46241, Republic of Korea}
\affiliation{Department of Earth Sciences, Pusan National University, Busan 46241, Republic of Korea}

\author[0000-0002-6925-4821]{Dohyeong~Kim}
\affiliation{Institute for Future Earth (IFE), Pusan National University, Busan 46241, Republic of Korea}
\affiliation{Department of Earth Sciences, Pusan National University, Busan 46241, Republic of Korea}
\affiliation{Department of Earth Science Education, Pusan National University, Busan 46241, Republic of Korea}

\author{Seonghun~Lim}
\affiliation{Department of Earth Sciences, Pusan National University, Busan 46241, Republic of Korea}
\affiliation{Gyeongsangnamdo Science Education Center, 178-35, Jinui-ro, Jinseong-myeon, Jinju-si, Gyeongsangnam-do, Republic of Korea}

\author[0000-0002-8537-6714]{Myungshin~Im}
\affiliation{SNU Astronomy Research Center, Department of Physics and Astronomy, Seoul National University, Gwanak-gu, Seoul, 08826, Republic of Korea}
\affiliation{Astronomy Program, Department of Physics and Astronomy, Seoul National University, Gwanak-gu, Seoul 08826, Republic of Korea}

\author[0000-0003-4422-6426]{Hyeonho~Choi}
\affiliation{SNU Astronomy Research Center, Department of Physics and Astronomy, Seoul National University, Gwanak-gu, Seoul, 08826, Republic of Korea}
\affiliation{Astronomy Program, Department of Physics and Astronomy, Seoul National University, Gwanak-gu, Seoul 08826, Republic of Korea}

\author{Jaemin~Park}
\affiliation{Department of Earth Science Education, Pusan National University, Busan 46241, Republic of Korea}

\author{Keun-Hong~Park}
\affiliation{Miryang Arirang Astronomical Observatory, Miryang-Daegongwonro 86, Miryang-si 50420, Gyeongsangnam-do, Republic of Korea}
\affiliation{Miryang City Facilities Management Corporation, Sicheongro 26, Miryang-si 50420, Gyeongsangnam-do, Republic of Korea}

\author{Junyeong~Park}
\author[0009-0000-2412-2609]{Chaudhary~Muskaan}
\affiliation{Department of Earth Sciences, Pusan National University, Busan 46241, Republic of Korea}

\author{Donghyun~Kim}
\affiliation{Department of Earth Science Education, Pusan National University, Busan 46241, Republic of Korea}

\author[0009-0003-4018-9995]{Hayeong~Jeong}
\affiliation{Department of Earth Sciences, Pusan National University, Busan 46241, Republic of Korea}
\affiliation{Department of Physics, Pusan National University, Busan 46241, Republic of Korea}




\begin{abstract}
    We introduce a 0.7\,m telescope system at the Miryang Arirang Astronomical Observatory (MAAO), a public observatory in Miryang, Korea. System integration and a scheduling program enable the 0.7\,m telescope system to operate completely robotically during nighttime, eliminating the need for human intervention. Using the 0.7\,m telescope system, we obtain atmospheric extinction coefficients and the zero-point magnitudes by observing standard stars. As a result, we find that atmospheric extinctions are moderate but they can sometimes increase depending on the weather conditions. The measured $5\sigma$ limiting magnitudes reach down to $BVRI$$=$$19.4–19.6$\,AB mag for a point source with a total integrated time of $10$\,minutes under clear weather conditions, demonstrating comparable performance with other observational facilities operating under similar specifications and sky conditions. We expect that the newly established MAAO 0.7\,m telescope system will contribute significantly to the observational studies of astronomy. Particularly, with its capability for robotic observations, this system, although its primary duty is for public viewing, can be extensively used for the time-series observation of transients.
\end{abstract}



\section{Introduction} \label{sec1}
    The role of small telescopes with an aperture size of $\lesssim$1\,m is still crucial in modern time-domain astrophysics. With the advantages of a wide field of view (FOV), low price, and easy accessibility, small telescopes remarkably increased the number of discoveries and the light curves of transients (e.g., classical novae, supernovae (SNe), luminous blue variables, etc.) in various surveys such as the Palomar Transient Factory \citep{2009PASP..121.1395L}, the Catalina Real-Time Transient Survey \citep{2009ApJ...696..870D}, the All-Sky Automated Survey for Supernovae \citep{2014ApJ...788...48S}, the Asteroid Terrestrial-impact Last Alert System \citep{2019PASP..131a8002B,2018PASP..130f4505T}, the Zwicky Transient Facility \citep{2019PASP..131g8001G}, the Korea Microlensing Telescope Network \citep{2016JKAS...49...37K}, and the Intensive Monitoring Survey of Nearby Galaxies \citep{2019JKAS...52...11I}.

    Recently, transients at short timescales (e.g., less than a few days), referred to as fast transients have been detected. Fast transients includes SNe early light curves \citep{2017ApJ...845L..11H, 2022ApJ...933L..45H, 2022NatAs...6..568N, 2023ApJ...949...33L}, M-dwarf flares \citep{2019ApJ...876..115S, 2020ApJ...892..144R}, optical afterglow of gamma-ray bursts \citep{2004RvMP...76.1143P, 2022MNRAS.512.2337D, 2022MNRAS.513.2777K}, and kilonovae \citep{2018NatAs...2..751L,2020NatAs...4...77J}. These fast transients allow us to understand their progenitor systems, explosion mechanisms, and interactions with their ambient environment, making the observation of fast transients important. However, due to short duration time of fast transients \citep[e.g. $\sim$ a sub-hour to days;][]{2020MNRAS.491.5852A, 2022AJ....163...95S}, many aspects of their nature remain uncovered.

    To successfully detect these fast transients, observations should be performed immediately when the transients occur, using a system that operates without human intervention, known as robotic observation. Robotic telescopes can perform scheduled observations based on the programmed observing plans, which reduces costs and uses nighttime fully. It is not a surprise that many small telescopes have already been operated robotically \citep{2015MNRAS.454.4316H,2015JKAS...48..207I,2016PASP..128j5005K,2019PASP..131f8003B,2022PASP..134c5003L}. These telescopes can respond fast to transient alerts, giving us the chance to perform imaging and spectroscopic follow-up observations using other telescope networks (e.g., \citealt{2021JKAS...54...89I}).

    The 0.7\,m telescope system was installed in 2020 May in the 7.7\,m diameter round dome at MAAO in Miryang, Korea\,($35^\circ30\farcm08\farcs6$N, $128^\circ45\farcm40\farcs0$E, $95$\,m). While MAAO primarily serves as a science museum for public education, it can also function as an astronomical research facility. 
    
    This paper is organized as follows. We present the characteristics of the system in Section~\ref{sec2}. In Section~\ref{sec3}, we describe the schematic of the system integration and the process of the robotic observation based on the scheduler. The system performances, such as point spread function (PSF) shape variation, standard flux calibration, and limiting magnitudes, are presented in Section~\ref{sec4}. Sections~\ref{sec5} and Section~\ref{sec6} discusses current research topics and future prospects, and in Section~\ref{sec7}, a summary is provided. In this work, all magnitudes are in the AB system unless otherwise specified.

\section{System Characteristics} \label{sec2}
        
    \subsection{Telescope and Mount} \label{sec2.1}
        The optical system of the 0.7\,m telescope is designed as a Corrected Dall-Kirkham system (CDK700) with a focal ratio of 6.5, which is produced by PlaneWave Instruments. It consists of three main components: an ellipsoidal primary mirror (700\,mm in diameter), a spherical secondary mirror (312.4\,mm in diameter), and an additional lens group to prevent image distortions such as coma, off-axis astigmatism, and field curvature. Both mirrors are made from fused silica, a material known for its excellent optical properties. Figure~\ref{fig1} shows the telescope mounted on an alt-azimuth mount system and equipped with dual Nasmyth foci, allowing for visual and CCD observations. 
        
        \begin{figure}[t!]
            \centering
            \includegraphics[width=\columnwidth]{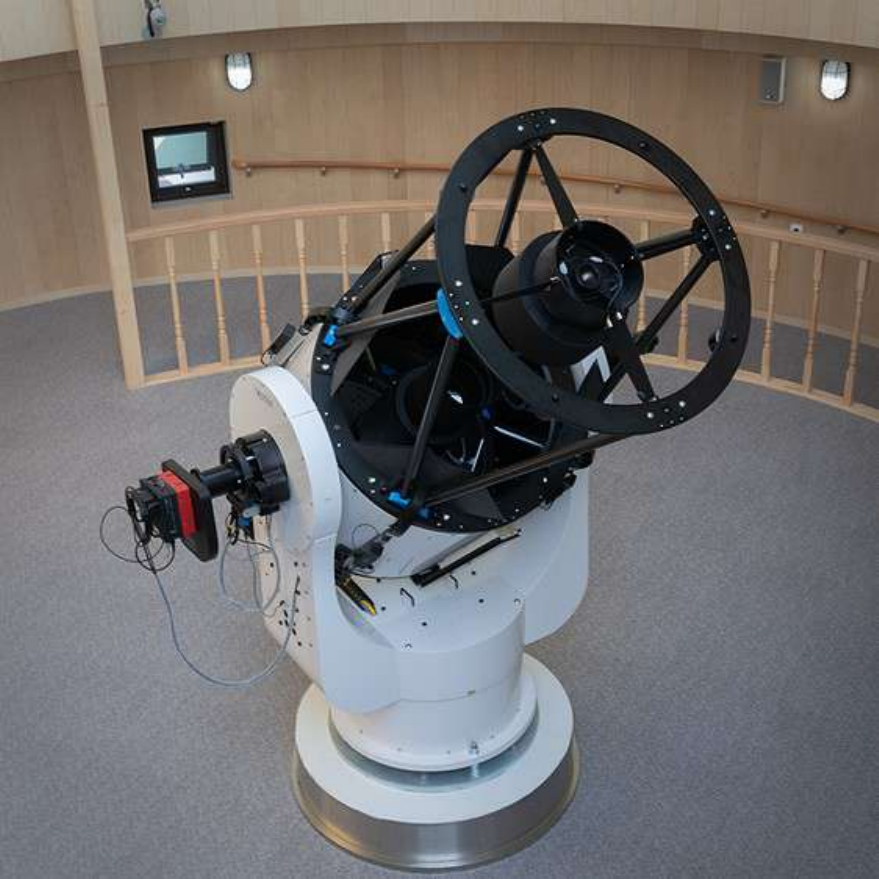}
            \caption{The 0.7\,m telescope at MAAO.}
        \end{figure} 
        \label{fig1}
        
        For the mount performance, slewing speed can reach up to 50\,deg\,s$^{-1}$ at maximum. The pointing accuracy is 3$\farcs$6 (rms) from the pointing model recently built with 102 stars at altitudes ranging from 20 to 75\,deg. The tracking accuracy is also measured using a similar method in previous studies (e.g., \citealt{2015MNRAS.454.4316H, 2015JKAS...48..207I, 2020PASP..132l5001Z}) by obtaining 60 images of each 1\,minute single exposure at an altitude of $\sim$50 deg. The angular distance between the center coordinates of a star in each image is accumulated, resulting in 0$\farcs$06\,min$^{-1}$, which is in agreement with the manufacturer's specifications.
        
    \subsection{Detector} \label{sec2.2}
        On the 0.7\,m telescope, an SBIG STX-16803 CCD camera is currently attached. The camera uses a KODAK's KAF-16803 chip with a $4096\times4096$ pixel array in each pixel size of $\rm 9\,\mu m \times 9\,\mu m$. We investigated the characteristics of this CCD in the dome with calibration images.
        
        \subsubsection{Pixel Scale, Gain, Readout Noise, and Flat Image}\label{sec2.2.1}
        
            The pixel scale ($P$) is given by
            
            \begin{equation} \label{eq1}
                P = \frac{206265 \times \mu}{1000 \times f}~~\rm (\arcsec \, pixel^{-1}),
            \end{equation}  
            
            where $\mu$ is the CCD pixel size in the unit of $\mu$m and $f$ is the focal length in the unit of mm. Considering the focal length of 4540\,mm of the 0.7\,m telescope, $P$ is calculated as $0\farcs41$\,pixel$^{-1}$ from Equation~\ref{eq1}. Then a field of view is $27.9\arcmin \times 27.9\arcmin$.
            
            In addition, we obtain a series of bias and sky-flat images. Taking the average values of the 2D arrays of each image ($\rm \bar B_{1}$, $\rm \bar B_{2}$, $\rm \bar F_{1}$, and $\rm \bar F_{2}$), the gain can be described as
            
            \begin{equation} \label{eq2}
                \mathrm{Gain}=\frac{(\bar F_{1}+\bar F_{2})-(\bar B_{1}+\bar B_{2})}{\sigma^{2}_{F_{1}-F_{2}}-\sigma^{2}_{B_{1}-B_{2}}},
            \end{equation}
            
            where $\sigma_{F_{1} - F_{2}}$ and $\sigma_{B_{1} - B_{2}}$ are the standard deviation of the difference images of two bias ($B_{1} - B_{2}$) and flat-field images ($F_{1} - F_{2}$). Flat images are flat-corrected using the master flat. The readout noise is calculated as
            \begin{equation} \label{eq3}
                \mathrm{Readout~Noise} = \frac{\mathrm{Gain} \cdot \sigma_{B_{1} - B_{2}}}{\sqrt{2}}.
            \end{equation}           
            Using Equations~\ref{eq2} and~\ref{eq3}, we obtain the gain and the readout noise as $1.30\pm0.04\,\rm e^{-}\,ADU^{-1}$ and $12.62\pm0.25\,\rm e^{-}$, which are consistent with the values in the specification from the manufacturer within the error.

            Furthermore, during the comparison between sky-flats and dome-flats by dividing them, the displacement of the donut pattern ($\sim$20–30\,pixels) is found, but this issue arises among sky-flats or dome-flats and sometimes does not occur. This offset issue is under investigation to resolve.

        \subsubsection{Residual Signal} \label{sec2.2.2}
\begin{figure*}[t!]
                \centering
                \includegraphics[width=0.7\textwidth]{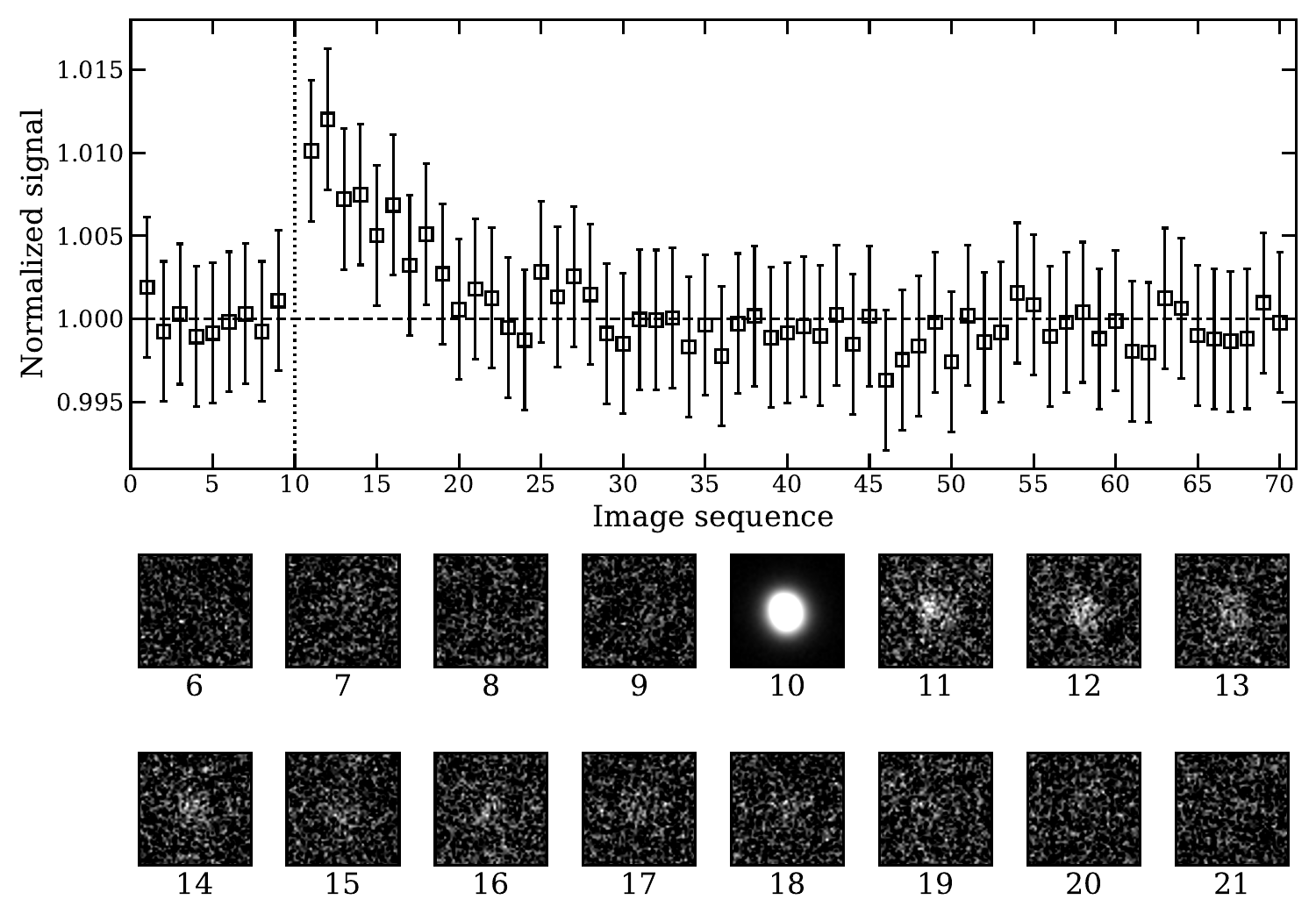}
                \caption{The signal variation with the image sequences for the residual signal. The error bars are photometric uncertainties of each data point. The vertical dotted line is the image sequence of the star, meaning that Group A and Group B are on the left and right sides of the vertical line, respectively. The horizontal dashed line is the mean value of Group A's signal. Thumbnails from the 6th to the 21st image are presented in a square box with a width of 50 pixels, which are Gaussian smoothed in a log scale.}
                \label{fig2}
            \end{figure*}

            We investigate the residual signal remaining at the position of a bright source after exposure. A bright star is exposed (300\,s) and a series of 30\,s dark frames are obtained before (9 frames; Group A) and after (60 frames; Group B) the star's image. Figure~\ref{fig2} shows the signal variation with the image sequence within a 10\,pixel diameter circular aperture (assuming $4.1\farcs$ seeing) at the star's position in the pixel coordinates. Each data point is normalized with the mean value of Group A's signal, and the star's signal (Image sequence = 10) is not shown due to its high signal value (45.88). We find the residual signals in Group B shown in the image thumbnails in Figure~\ref{fig2}. The signal of the first dark frame in Group B (Image sequence = 11) is 1\% higher than the mean value of Group A's signal, resulting in a magnitude difference of 0.011\,mag. The effect of the residual signal disappears within 3 times the standard deviation of Group A's signal after 8 readouts. Therefore, observers have to consider the residual signal for the photometry of the faint objects.
            
        \subsubsection{Linearity} \label{sec2.2.3}
\begin{figure}[t!]
                \centering
                \includegraphics[width=\columnwidth]{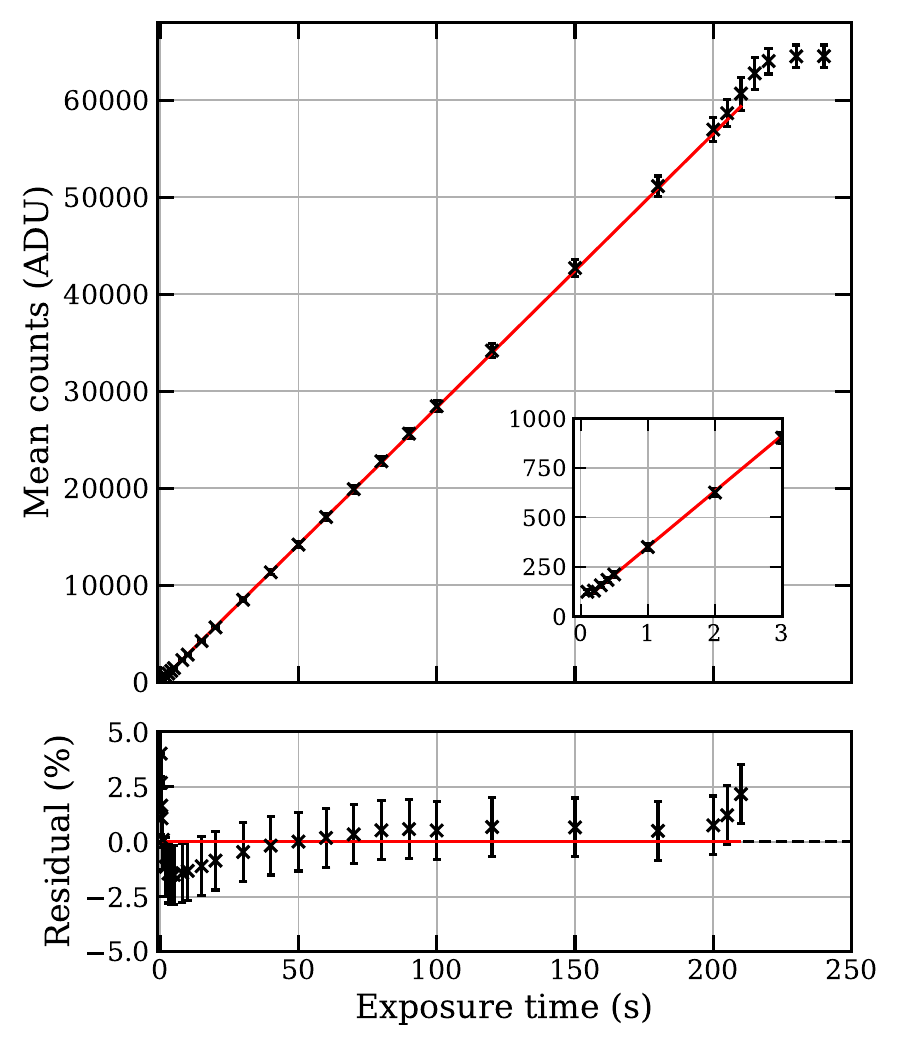}
                \caption{The variation of the mean values of the dome-flat images is presented. The linear fit is overplotted in a red solid line from the exposure time of 0.2–210\,s. The inset plot shows the linear relation for shorter exposures. The residual plot is also provided in the bottom panel.}
                \label{fig3}
            \end{figure}

            It is important to check if the input charges can be converted into the output electric signal with a simple linear relation. A series of dome-flat images taken with $V$-band are obtained by progressively increasing exposure time while the dome-flat lamps are turned on and the telescope is directed toward the dome-flat panel. Figure~\ref{fig3} shows a variation of mean counts of the dome-flat images from $0.1$ to $240$\,s exposure. The pixels started to be saturated from $215$\,s exposure. The mean counts are well fitted with the linear relation up to $\sim61,000$\,ADU within $2.5$\,\% and $\sim57,000$\,ADU within $1$\,\% ($\chi^{2}_{\nu} = 0.16$). Even in the exposure time shorter than $1$\,s, the linearity can be found $\gtrsim100$\,ADU within $5$\,\%. During this process, any feature of the shutter pattern for the short-exposure images is not found.

        \subsubsection{Dark Current} \label{sec2.2.4}
\begin{figure}[t!]
                \centering
                \includegraphics[width=\columnwidth]{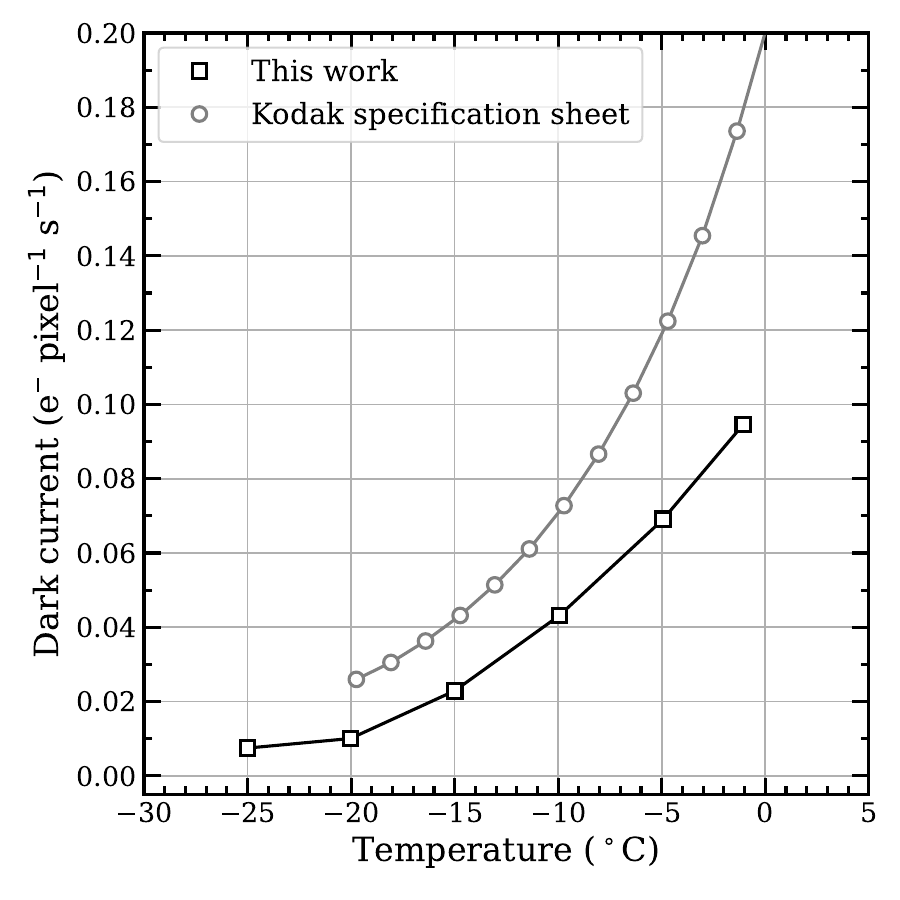}
                \caption{A variation of the dark current with the CCD temperature. The data points indicated by gray circles are obtained from the Kodak KAF-16803 Image Sensor Device Performance Specification document (\url{https://www. onsemi.com/products/sensors/image-sensors/KAF-16801}).}
                \label{fig4}
            \end{figure}

            We also investigate a variation of the dark current resulting from the thermal electron with the CCD cooling temperatures from $-30\,^{\circ}$C to $-1.1\,^{\circ}$C. At $-20\,^{\circ}$C, the dark current is $0.01\,\rm e^{-}\,pixel^{-1}\,s^{-1}$ as presented in Figure~\ref{fig4}. The mean values from the bias-subtracted and combined dark frames ($180$\,s exposure) are taken. The dark frame obtained at the lowest temperature ($-30\,^{\circ}$C) is used as a reference to remove any other signals not dependent on the CCD temperature by subtracting those taken at the high temperature.

    \subsection{Filter System and Throughput} \label{sec2.3}
\begin{figure}[t!]
            \centering
            \includegraphics[width=\columnwidth]{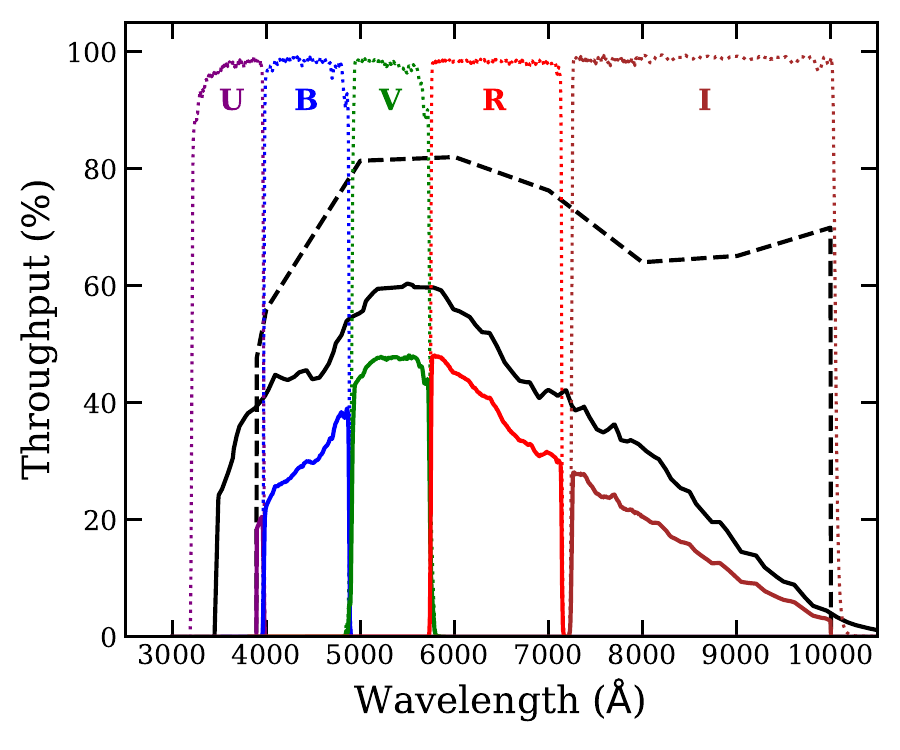}
            \caption{Transmission curves of the broadband filter system at MAAO. The optics, CCD QE, and each filter are plotted with black dashed, solid, and dotted lines, respectively. The filter transmission is plotted before and after considering CCD QE and optics (colored solid lines). Note that the $U$-band throughput is not complete since the transmission of the optics and CCD QE is not provided in a wavelength range $<$3900 and $<$3490\,{\rm \AA{}}.}
            \label{fig5}
        \end{figure}

        Currently, the filter wheel SBIG-FW7 attached to the CCD camera has a set of 5 broadband (Bessell $UBVRI$) and 3 narrow band filters (H$\alpha$, \ion{O}{3}, and \ion{S}{2}) manufactured by Chroma Technology, Co. with a size of $50$\,mm$\times$$50$\,mm. The filter transmission curve altogether with the quantum efficiency (QE) of CCD and the telescope optics throughput is provided in Figure~\ref{fig5}. For CCD, the QE peaks at $60$\,\% at $5500$\,{\rm \AA{}}. For optics throughput, the corrector lenses with the standard anti-reflection coatings and the first surface mirrors with the enhanced Aluminium reflective coatings are considered, showing the peak ($82$\,\%) at $6000$\,{\rm \AA{}}. 

        In summary, the overall characteristics of the CCD camera on the MAAO 0.7\,m telescope are presented in Table~\ref{table1}.
        
        \begin{table}[t!]
            \centering
            \caption{Specifications of the current CCD camera (SBIG STX-16803).}
            \label{table1}
            \begin{tabular}{lc}
            \hline
            Properties & Values \\
            \hline
            Sensor chip & KAF-16803 (36.8\,mm $\times$ 36.8\,mm)\\
            Pixel size & 9\,$\mu$m$\times$9\,$\mu$m\\
            Pixel arrays & 4,096$\times$4,096 \\
            Readout time\tablenotemark{a} & 9 s \\
            QE peak & 60\,\% \\
            Pixel scale & 0$\farcs$41\,pixel$^{-1}$ \\
            Field of view & 27$\farcm$9 $\times$ 27$\farcm$9 \\
            Gain & 1.30$\pm$0.04 e$^{-}$\,ADU$^{-1}$ \\
            Readout noise & 12.62$\pm$0.25 e$^{-}$ \\
            Linearity & 350$-$57,000\,ADU (within 1\,\%)\\
            Full well\tablenotemark{a} & 100,000 e$^{-}$ \\
            Dark current & 0.01 e$^{-}$\,pixel$^{-1}$\,s$^{-1}$ at $-20^\circ$C\\
            \hline
            \end{tabular}
            \tablecomments{$^{a}$ The manufacturer's data are adopted.}
        \end{table}

\section{System Integration and Robotic Observation} \label{sec:software and operation} \label{sec3}
    \subsection{System Integration} \label{sec3.1}
\begin{figure*}[t!]
        \centering
        \includegraphics[width=0.85\textwidth]{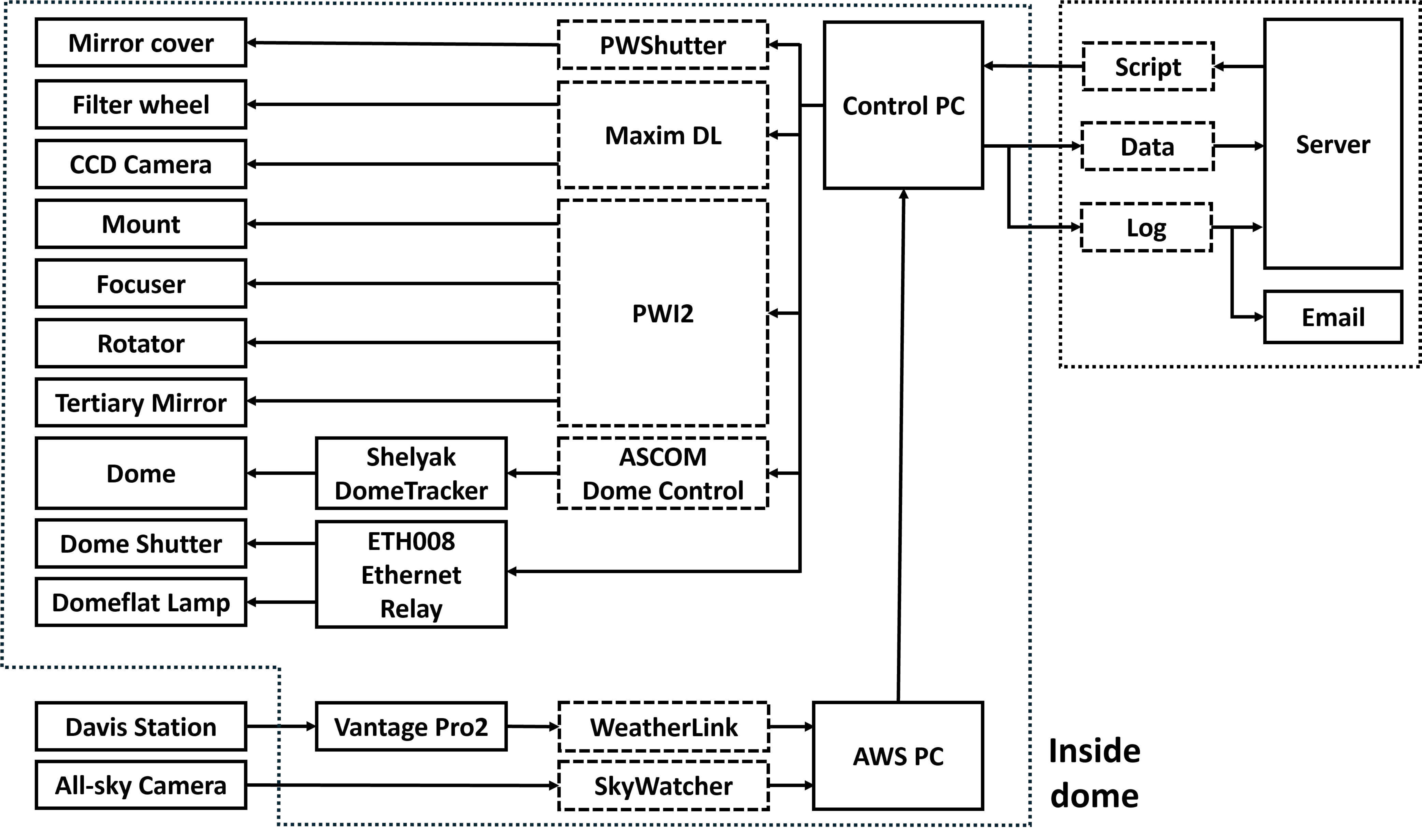}
        \caption{A schematic of the system integration at MAAO. Solid boxes represent hardware, while dashed lines indicate computer files and software components, including scripts, data, and logs.}
        \label{fig6}
    \end{figure*}

        Two personal computers (PCs) control the system in a 64-bit Windows operating system environment. One is the main control PC connected to the 0.7\,m telescope system and the other is the automatic weather station (AWS) PC. Figure~\ref{fig6} summarizes the system configuration at MAAO.
        
        \subsubsection{Telescope System} \label{sec3.1.1}
            The 0.7\,m telescope system is operated by a series of commercial software in the main control PC such as \textsc{PlaneWave Interface 2} (\textsc{PWI2}), \textsc{PWShutter}, and \textsc{MaxIm DL 6.26}. We integrated all the instruments using \textsc{Python} scripts, as demonstrated below. 
            
            The mount, the derotator, the tertiary mirror, and the focuser are controlled by \textsc{PWI2} over TCP/IP communication on a local network. \textsc{PWI2} command is also supported by a \textsc{Python} script provided by PlaneWave Instruments\footnote{\url{https://planewave.com/files/software/PWI2/pwi2control.py}}. The mirror cover control software, \textsc{PWShutter}, is connected in the same manner referring to the sample code provided in the \textsc{PWShutter} user manual.
            
            For the CCD camera and the filter wheel control, \textsc{MaxIm DL 6.26}\footnote{\url{https://diffractionlimited.com/product/maxim-dl/}} is utilized using \textsc{AStronomy Common Object Model (ASCOM)}\footnote{\url{https://ascom-standards.org/}} standard driver, which provides universal interfaces for many astronomical devices. The ASCOM properties and methods in \textsc{MaxIm~DL} can be translated into \textsc{Python} language via \texttt{pywin32}\footnote{\url{https://github.com/mhammond/pywin32}} module.
            
            The dome is equipped with a Shelyak Dome Tracker in a serial connection for the \textsc{ASCOM dome control}. For the dome shutter and dome-flat lamps, they are independently controlled by an ethernet relay via TCP/IP communication using another IP address.
            
        \subsubsection{Weather Monitoring System} \label{sec3.1.2}
            There are two components to the weather monitoring system: the Davis weather station for meteorological data and the all-sky camera for measuring cloud coverage.
            
            The Davis weather station obtains temperature, humidity, wind speed/direction, and so on. This equipment is connected to the AWS PC via a Vantage Pro2 device on a serial port connection. We upload the weather data in real-time on the global network\footnote{\url{https://www.weatherlink.com/}} via \textsc{WeatherLink} software. Then we retrieve this data via a \textsc{Python} script using \texttt{WeatherLink APIv2}\footnote{\url{https://weatherlink.github.io/v2-api/}} on the main control PC to give constraints on whether to open or close the dome.

            Using the all-sky camera images, we detect the presence of clouds with a \textsc{Python} module of \texttt{cloudynight}\footnote{\url{https://github.com/mommermi/cloudynight}} \citep{2020AJ....159..178M}. The \textsc{Python} module uses a machine-learning model of \texttt{lightGBM} \citep{Ke2017}. The training and test dataset will be addressed in the future work (Lim et al. in preparation). For in-person observations, observers can identify clouds in images visually.

            The dome remains open when all of the following conditions are met: outside humidity~$<$95\,\%, outside temperature~$>$$-10$\,$^{\circ}$C, wind speed~$<$15\, $\rm m\,s^{-1}$, no cloud detection, and no detection of lightning within a radius $<$$16\, \rm km$ around MAAO.\footnote{We adapted the radius in the Lick observatory (\url{https://mthamilton.ucolick.org/techdocs/telescopes/Nickel/limits_weather/}), which is related to the standard lightning alert radius \citep{2022BAMS..103E.548D}).}
             
    \subsection{Robotic Observation} \label{sec3.2}
\begin{figure}[t!]
            \centering
            \includegraphics[width=0.9\columnwidth]{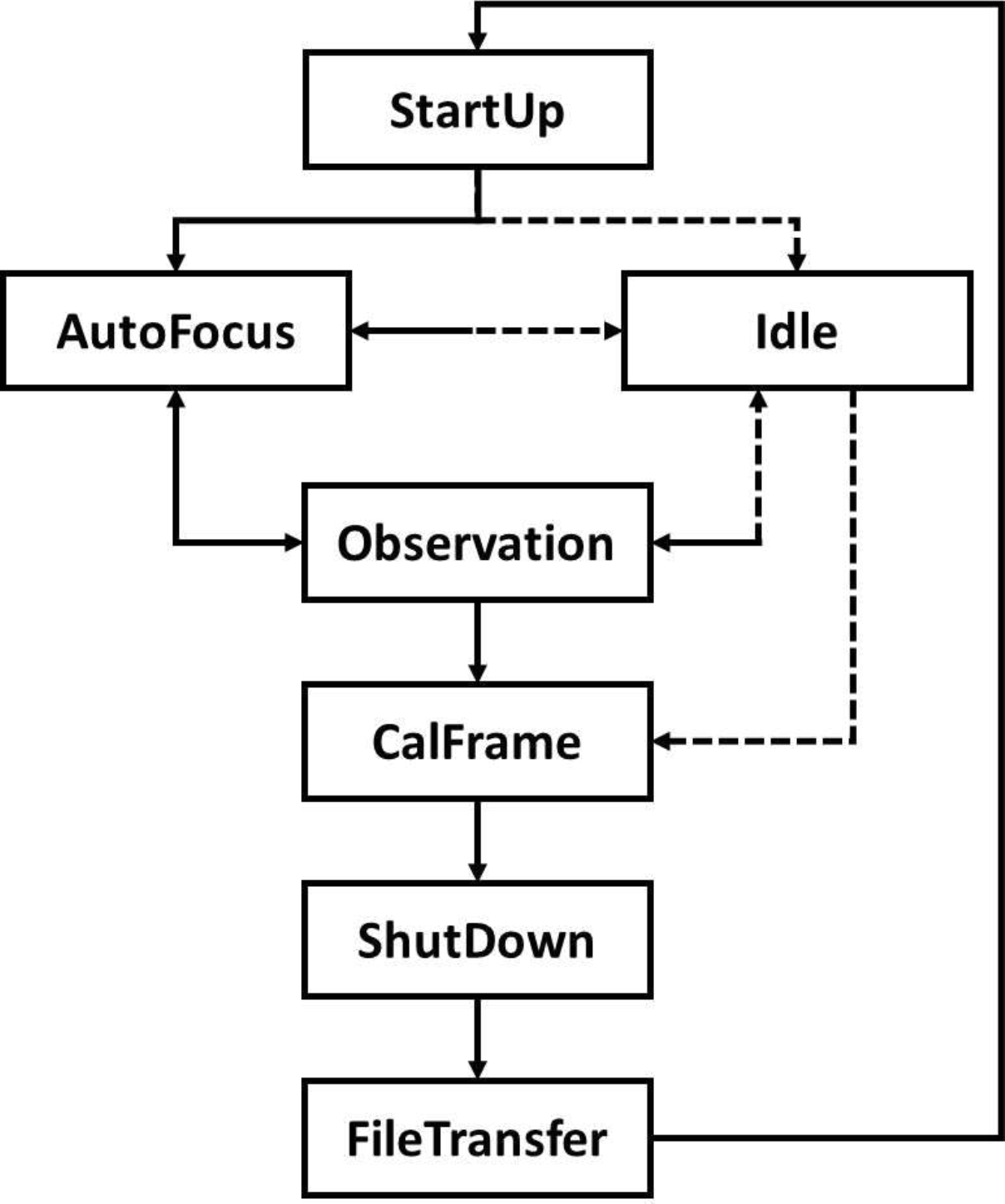}
            \caption{A flowchart for the scheduler at MAAO and each session. Solid arrows show the path to the next step when the weather is safe, while dashed arrows represent the path under unsafe weather conditions.}
            \label{fig7}
        \end{figure}

        An one-day run of the robotic observation involves 7 sessions: \texttt{StartUp}, \texttt{Observation}, \texttt{AutoFocus}, \texttt{Idle}, \texttt{CalFrame}, \texttt{ShutDown}, and \texttt{FileTransfer}, and they are controlled by a scheduler. Figure~\ref{fig7} illustrates a sequence of these sessions. 

        (i) \texttt{StartUp} session can start when the solar altitude is below $-18$\,deg and when the public event is finished (after $9$\,PM/$10$\,PM in winter/summer). This session includes software initialization, azimuth synchronization of the dome and scope, dome opening, tertiary mirror position switching for CCD imaging, and CCD chip cooling. When the weather condition is poor, the scope enters \texttt{Idle} session. 

        (ii) \texttt{Observation} session performs observations based on scripts. The observation script, an ASCII file, must contain the following columns: object name (\texttt{NAME}), R.A. in J2000 (\texttt{RA\_J2000}), decl. in J2000 (\texttt{DEC\_J2000}), rise and set times above a specific altitude in local time (\texttt{RISE(LT)}, \texttt{SET(LT)}), transit time (\texttt{TRANSIT(LT)}), filter (\texttt{FILTER}), single exposure time (\texttt{EXPTIME}), CCD binning (\texttt{BINNING}), the number of exposures to repeat (\texttt{COUNTS}), and observation starting time in local time (\texttt{OBSTIME}). 

        (iii) \texttt{AutoFocus} session uses the default auto-focusing function supported in \textsc{PWI2}. In the \texttt{AutoFocus} session, an optimal focus value can be determined by using a relationship between PSF sizes of a point source and focus values in a series of short exposure images. Users have the option to perform \texttt{AutoFocus} manually by generating an \texttt{Observation} session with ``AutoFocus'' in the \texttt{NAME} column. An initial auto-focusing procedure is performed every \texttt{StartUp} session.

        (iv) \texttt{CalFrame} session takes calibration frames (bias, dark, and dome-flats) before \texttt{ShutDown} session. For dome-flats, after the dome closed, the scope points to the dome-flat panel (Azimuth $= 267$\,deg, altitude $= 51.5$\,deg), and the dome moves to the opposite position (Azimuth $= 88.3$\,deg).

        (v) \texttt{Idle} session moves the scope to the parking position, disables tracking, and closes the dome. The observation resumes when the weather conditions have remained safe for 30\,minutes.

        (vi) \texttt{ShutDown} session is set to automatically initiate when the solar elevation exceeds $-18$\,deg. 

        (vii) \texttt{FileTransfer} session automatically sends observed data obtained during the nighttime using the file transfer protocol (FTP) to the server. The observation is summarized into an observation log and sent to the observer via email.

        During the execution, the scheduler checks the solar altitude, local time, and weather data every 60\,s to finish or halt the observation. An operation log is also generated after each command.

\section{System Performances} \label{sec4}
    We investigate PSF shape variations, atmospheric coefficients, zero-point magnitudes, and limiting magnitudes. Acquired raw data in this section is reduced with the standard image reduction and analysis facility (\textsc{IRAF}) routine using \textsc{PyRAF} package \citep{2012ascl.soft07011S} of bias, dark, and flat-field correction. Although it is known that observations at IR wavelengths can be affected by fringe patterns (e.g. Fig 13 in \citealt{2010JKAS...43...75I}), we find no significant fringe pattern even in $I$-band images, and therefore, no fringe pattern correction is applied. Astrometric solution is entered using \textsc{Astrometry.net} \citep{2010AJ....139.1782L}. The data in this work are reduced in this manner.
    
    \subsection{Optical Performance} \label{sec4.1}
\begin{figure*}[t!]
            \centering
            \includegraphics[width=0.45\textwidth]{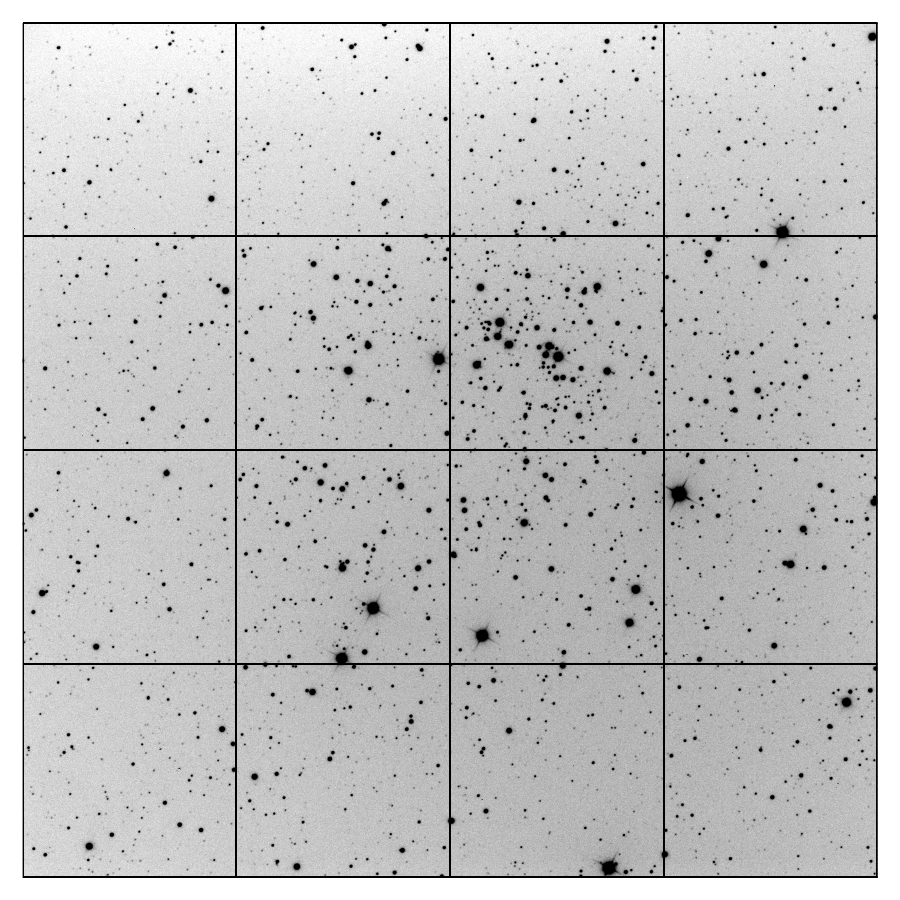}
            \hspace{0.01\textwidth}
            \includegraphics[width=0.45\textwidth]{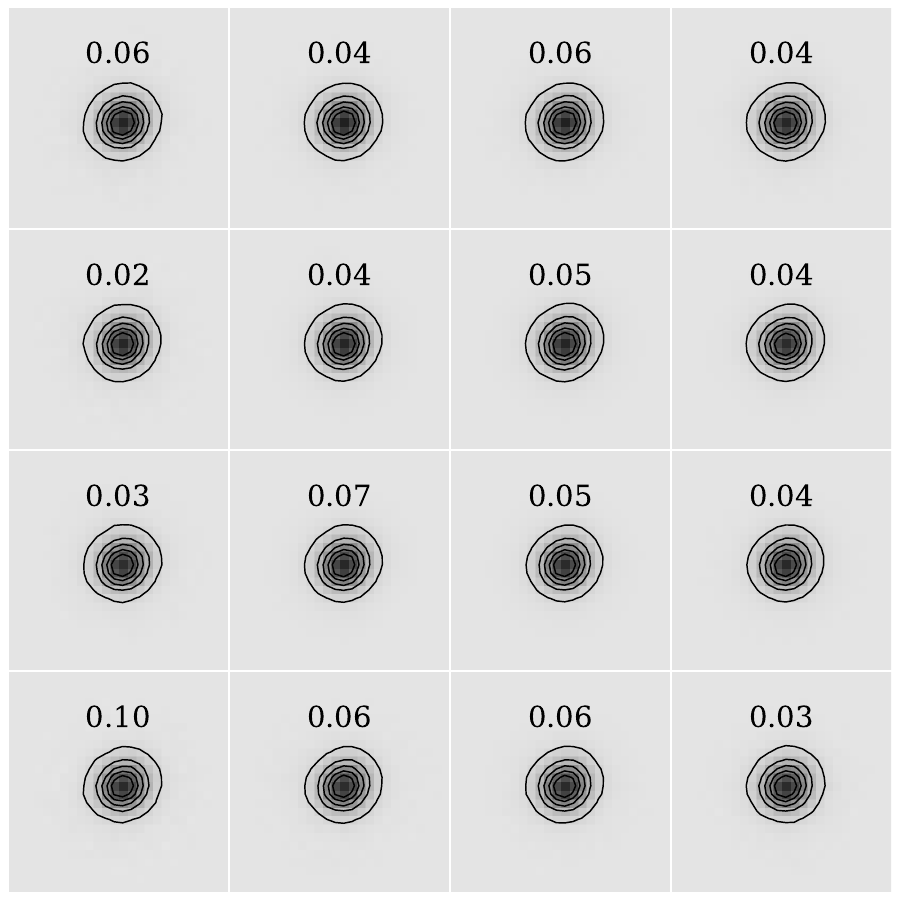}
            \caption{Left: The total FOV of the CCD camera for the $R$-band image divided into 16 sections. The mean full-width half maximum (FWHM) value in the total field is 2$\farcs$9 for the 60\,s exposure time. Right: PSF images built from stars in each division. The number presents the ellipticity of the PSF.}
            \label{fig8}
        \end{figure*}

        To explore the optical performance and tracking accuracy, we obtain the images with extending exposure times from $5$ to $600$\,s, including $10$, $30$, $60$, $120$, $180$, and $300$\,s. The left panel in Figure~\ref{fig8} shows the total FOV divided into 16 image sections a 60\,s exposure of an open cluster NGC 0884. The PSF model in each section is built using \textsc{PSFEx} \citep{2013ascl.soft01001B} from the detected stars with a high S/N ratio at least 20 but not saturated by setting the \textsc{PSFEx} parameters of \texttt{SAMPLE\_MINSN} = 20 and \texttt{SATUR\_LEVEL} = 50000, respectively. The right panel shows the PSF models produced in each image section with the ellipticity values (\texttt{ELLIPTICITY} from \textsc{SExtractor}, \citealt{1996A&AS..117..393B}). Though the PSF shape on the left-bottom side is slightly elongated ($0.1$ at maximum), we find uniform PSF shapes over the FOV. 

    \subsection{Photometric Calibration} \label{sec4.2}
        Photometric calibration is required because the photometric system depends on the characteristics of optics, imaging devices, and observing conditions. To measure the photometric parameters using data obtained at MAAO, we observed several standard star fields (SA 23 and SA 35) presented in \citet{2013AJ....146..131L}. The observations were performed in the clear sky, and the observation information is summarized in Table ~\ref{table2}. After data reduction, we measure the magnitudes of the standard stars using the SExtractor with the $14\arcsec$ diameter circular aperture to match with the same size aperture in \citet{2013AJ....146..131L}. We obtained the photometric parameters using the following equations:
 
        \begin{equation}
            b - B = z_{B} + k_{B}^{\prime}X + k_{B}^{\prime\prime}(B-V)X + C_{B}(B-V),
        \end{equation}  
        \begin{equation}
            v - V = z_{V} + k_{V}^{\prime}X + k_{V}^{\prime\prime}(B-V)X + C_{V}(B-V),
        \end{equation}  
        \begin{equation}
            r - R = z_{R} + k_{R}^{\prime}X + k_{R}^{\prime\prime}(V-R)X + C_{R}(V-R),
        \end{equation}  
        \begin{equation}
            i - I = z_{I} + k_{I}^{\prime}X + k_{I}^{\prime\prime}(V-I)X + C_{I}(V-I).
        \end{equation}    
        
        In these equations, the observed magnitudes are denoted as $b$, $v$, $r$, and $i$ and the standard magnitudes are $B$, $V$, $R$, and $I$ provided from \citet{2013AJ....146..131L}. Additionally, $z_{B}$, $z_{V}$, $z_{R}$, and $z_{I}$ correspond to the zero-points for the magnitudes producing a signal of 1\,e$^{-}$\,s$^{-1}$ in each band. $k_B^{\prime}$, $k_V^{\prime}$, $k_R^{\prime}$, and $k_I^{\prime}$ are the first-order atmospheric extinction coefficients and $k_B^{\prime\prime}$, $k_V^{\prime\prime}$, $k_R^{\prime\prime}$, and $k_I^{\prime\prime}$ are the second-order atmospheric extinction coefficients. Since the second-order extinction coefficients are small, they can be negligible. The color term coefficients are $C_{B}$, $C_{V}$, $C_{R}$, and $C_{I}$, and $X$ represents the airmass. Table~\ref{table3} presents the resultant of the calibration.
    
        \begin{table*}[t!]
            \centering
            \caption{Observation logs of standard stars (SA 23 and SA 35)}
            \label{table2}
            \begin{tabular}{ccccccc}
                \hline\hline
                Epoch & Filter & Date-Obs (UT) & MJD & Airmass & FWHM ($''$) & Exposures \\
                \hline
                $1$ & $B$ & 2023-02-20 10:38:37 & $59995.443$ & $1.06$ & $3.18$ & $60$\,s$\times3$ \\
                & $V$ & 2023-02-20 10:45:21 & $59995.448$ & $1.07$ & $3.03$ & \\
                \hline
                $2$ & $B$ & 2023-02-20 12:17:41 & $59995.512$ & $1.24$ & $3.69$ & $60$\,s$\times3$ \\
                & $V$ & 2023-02-20 12:21:26 & $59995.515$ & $1.25$ & $3.57$ & \\
                \hline
                $3$ & $B$ & 2023-02-20 13:38:13 & $59995.568$ & $1.57$ & $2.86$ & $60$\,s$\times3$\\
                & $V$ & 2023-02-20 13:42:22 & $59995.571$ & $1.59$ & $2.98$ & \\
                \hline
                $4$ & $B$ & 2023-02-20 14:17:47 & $59995.596$ & $1.84$ & $4.14$ & $60$\,s$\times3$\\
                & $V$ & 2023-02-20 14:21:51 & $59995.599$ & $1.87$ & $3.90$ & \\
                \hline
                $5$ & $B$ & 2023-02-20 14:39:39 & $59995.611$ & $2.03$ & $3.56$ & $60$\,s$\times3$\\
                & $V$ & 2023-02-20 14:44:40 & $59995.614$ & $2.09$ & $3.85$ & \\
                \hline
                \hline
                $1$ & $B$ & 2023-03-28 13:13:48 & $60031.551$ & $2.19$ & $3.48$ & $60$\,s$\times5$ \\
                & $V$ & 2023-03-28 13:19:54 & $60031.555$ & $2.12$ & $2.74$ & \\
                & $R$ & 2023-03-28 13:26:01 & $60031.560$ & $2.05$ & $3.71$ & \\
                & $I$ & 2023-03-28 13:32:08 & $60031.564$ & $1.99$ & $3.09$ & \\
                \hline
                $2$ & $B$ & 2023-03-28 13:57:20 & $60031.581$ & $1.77$ & $3.64$ & $60$\,s$\times5$ \\
                & $V$ & 2023-03-28 14:03:28 & $60031.586$ & $1.73$ & $3.66$ & \\
                & $R$ & 2023-03-28 14:09:35 & $60031.590$ & $1.68$ & $3.56$ & \\
                & $I$ & 2023-03-28 14:15:41 & $60031.594$ & $1.64$ & $3.51$ & \\
                \hline
                $3$ & $B$ & 2023-03-28 14:23:32 & $60031.600$ & $1.59$ & $4.31$ & $60$\,s$\times5$ \\
                & $V$ & 2023-03-28 14:29:39 & $60031.604$ & $1.56$ & $3.72$ & \\
                & $R$ & 2023-03-28 14:35:46 & $60031.608$ & $1.53$ & $3.34$ & \\
                & $I$ & 2023-03-28 14:41:52 & $60031.612$ & $1.49$ & $2.95$ & \\
                \hline
                $4$ & $B$ & 2023-03-28 14:55:39 & $60031.622$ & $1.43$ & $3.51$ & $60$\,s$\times5$ \\
                & $V$ & 2023-03-28 15:01:46 & $60031.626$ & $1.40$ & $3.56$ & \\
                & $R$ & 2023-03-28 15:07:53 & $60031.630$ & $1.38$ & $3.50$ & \\
                & $I$ & 2023-03-28 15:13:60 & $60031.635$ & $1.35$ & $3.09$ & \\
                \hline
                $5$ & $B$ & 2023-03-28 16:01:28 & $60031.668$ & $1.21$ & $3.81$ & $60$\,s$\times5$ \\
                & $V$ & 2023-03-28 16:07:34 & $60031.672$ & $1.19$ & $3.29$ & \\
                & $R$ & 2023-03-28 16:13:41 & $60031.676$ & $1.18$ & $2.75$ & \\
                & $I$ & 2023-03-28 16:19:48 & $60031.680$ & $1.16$ & $2.84$ & \\
                \hline
                \hline
                $1$ & $B$ & 2024-03-20 13:40:44 & $60389.570$ & $2.22$ & $4.42$ & $120$\,s$\times5$ \\
                & $V$ & 2024-03-20 13:51:50 & $60389.578$ & $2.09$ & $4.15$ & \\
                & $R$ & 2024-03-20 14:02:56 & $60389.582$ & $2.02$ & $4.07$ & \\
                & $I$ & 2024-03-20 14:14:02 & $60389.590$ & $1.91$ & $4.04$ & \\
                \hline
                $2$ & $B$ & 2024-03-20 14:29:45 & $60389.604$ & $1.75$ & $4.17$ & $120$\,s$\times5$ \\
                & $V$ & 2024-03-20 14:40:51 & $60389.612$ & $1.67$ & $3.52$ & \\
                & $R$ & 2024-03-20 14:51:56 & $60389.616$ & $1.63$ & $3.36$ & \\
                & $I$ & 2024-03-20 15:03:02 & $60389.624$ & $1.56$ & $3.53$ & \\
                \hline
                $3$ & $B$ & 2024-03-20 15:23:52 & $60389.642$ & $1.43$ & $3.96$ & $120$\,s$\times5$ \\
                & $V$ & 2024-03-20 15:34:58 & $60389.649$ & $1.38$ & $3.89$ & \\
                & $R$ & 2024-03-20 15:46:04 & $60389.654$ & $1.36$ & $3.64$ & \\
                & $I$ & 2024-03-20 15:57:10 & $60389.662$ & $1.32$ & $2.41$ & \\
                \hline
                $4$ & $B$ & 2024-03-20 16:18:30 & $60389.680$ & $1.24$ & $3.83$ & $120$\,s$\times5$ \\
                & $V$ & 2024-03-20 16:29:36 & $60389.687$ & $1.21$ & $3.45$ & \\
                & $R$ & 2024-03-20 16:40:42 & $60389.692$ & $1.19$ & $2.90$ & \\
                & $I$ & 2024-03-20 16:51:48 & $60389.700$ & $1.17$ & $2.63$ & \\
                \hline
            \end{tabular}
        \end{table*}
        
        \begin{table*}[t!]
        \centering
        \caption{The resultants of standard calibration.}
        \label{table3}
        \begin{tabular}{ccccccc}
        \hline\hline
        Filter & $k'$ & $C$ & $k''$ & $z$ & $N_{\rm star}$ & $\sigma_{\rm rms}$ \\
        \hline
            $B$ & $0.477\pm0.032$ & $0.011\pm0.070$ & $-0.070\pm0.047$ & $-21.632\pm0.049$ & $11$ & $0.028$\\
            $V$ & $0.328\pm0.013$ & $0.102\pm0.027$ & $-0.016\pm0.017$ & $-21.912\pm0.021$ & $14$ & $0.021$\\
        \hline
        \hline
            $B$ & $1.375\pm0.086$ & $-0.182\pm0.237$ & $-0.038\pm0.138$ & $-21.911\pm0.148$ & $7$ & $0.039$ \\
            $V$ & $1.068\pm0.041$ & $0.197\pm0.099$ & $-0.090\pm0.057$ & $-22.181\pm0.071$ & $7$ & $0.016$ \\
            $R$ & $0.752\pm0.028$ & $0.012\pm0.121$ & $0.084\pm0.069$ & $-21.861\pm0.049$ & $7$ & $0.010$ \\
            $I$ & $0.642\pm0.088$ & $0.192\pm0.118$ & $-0.100\pm0.067$ & $-21.081\pm0.088$ & $7$ & $0.019$ \\
        \hline
        \hline
            $B$ & $0.560\pm0.036$ & $-0.163\pm0.086$ & $0.031\pm0.053$ & $-21.777\pm0.059$ & $7$ & $0.019$ \\
            $V$ & $0.311\pm0.020$ & $-0.007\pm0.045$ & $0.072\pm0.028$ & $-21.960\pm0.032$ & $7$ & $0.011$\\
            $R$ & $0.221\pm0.027$ & $0.024\pm0.107$ & $0.110\pm0.066$ & $-21.865\pm0.044$ & $7$ & $0.012$\\
            $I$ & $0.120\pm0.022$ & $-0.014\pm0.048$ & $0.036\pm0.029$ & $-20.912\pm0.037$ & $7$ & $0.009$\\
        \hline
        \end{tabular}
        \tablecomments{$N_{\rm star}$ is the number of stars used in the calibration process.}
        \end{table*}
    
\begin{figure*}[ht]
            \centering
            \begin{tabular}{cc}
                \includegraphics[width=0.32\textwidth]{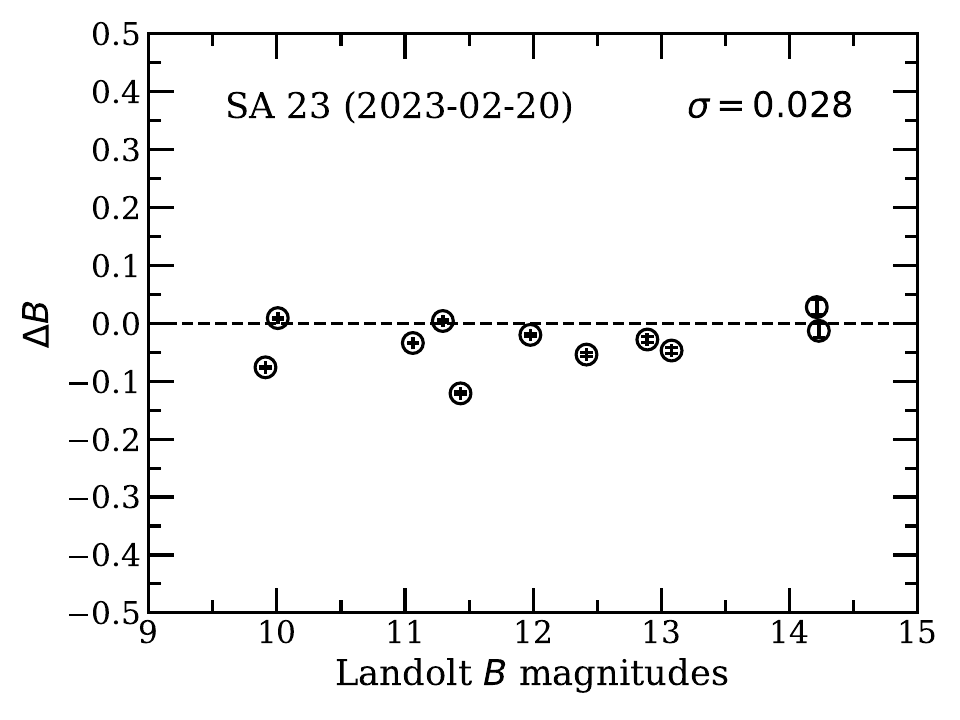} & \includegraphics[width=0.32\textwidth]{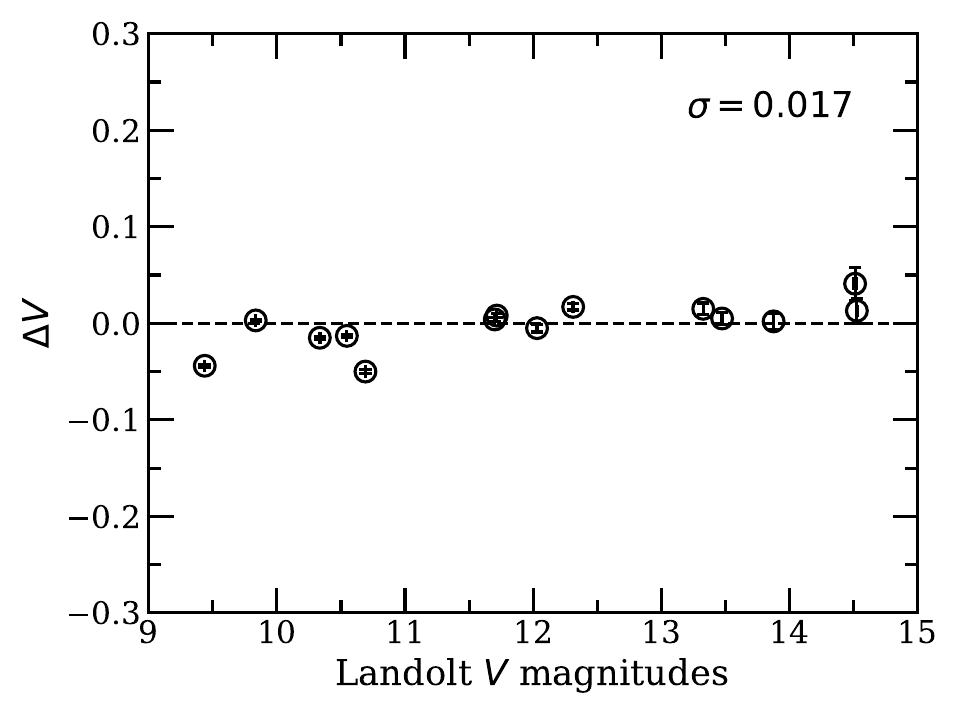} \\
                \includegraphics[width=0.32\textwidth]{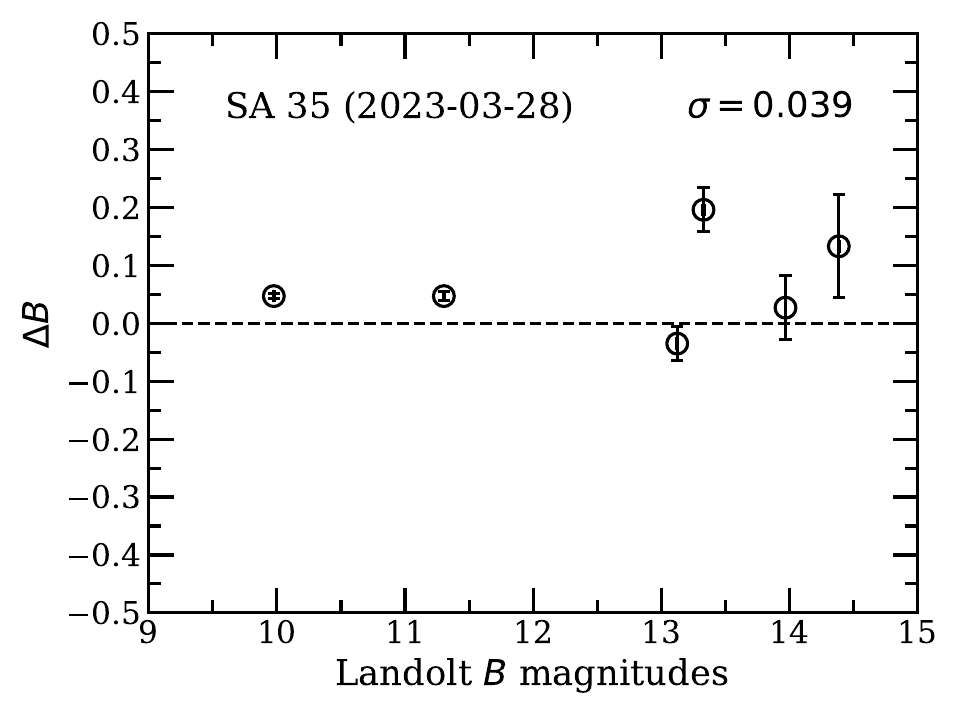} & \includegraphics[width=0.32\textwidth]{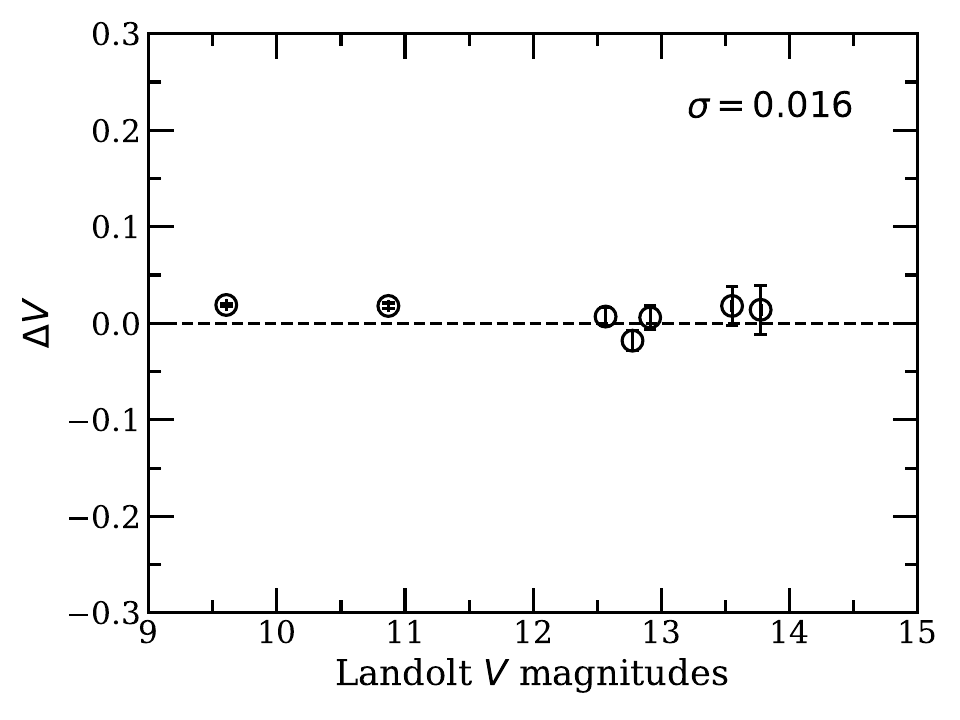} \\
                \includegraphics[width=0.32\textwidth]{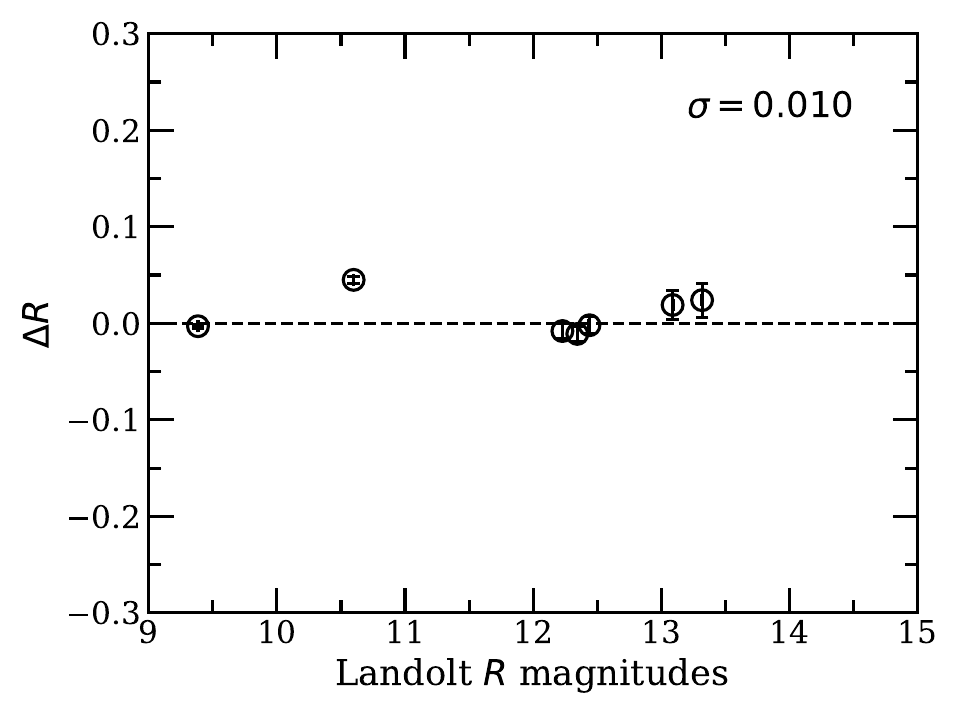} & \includegraphics[width=0.32\textwidth]{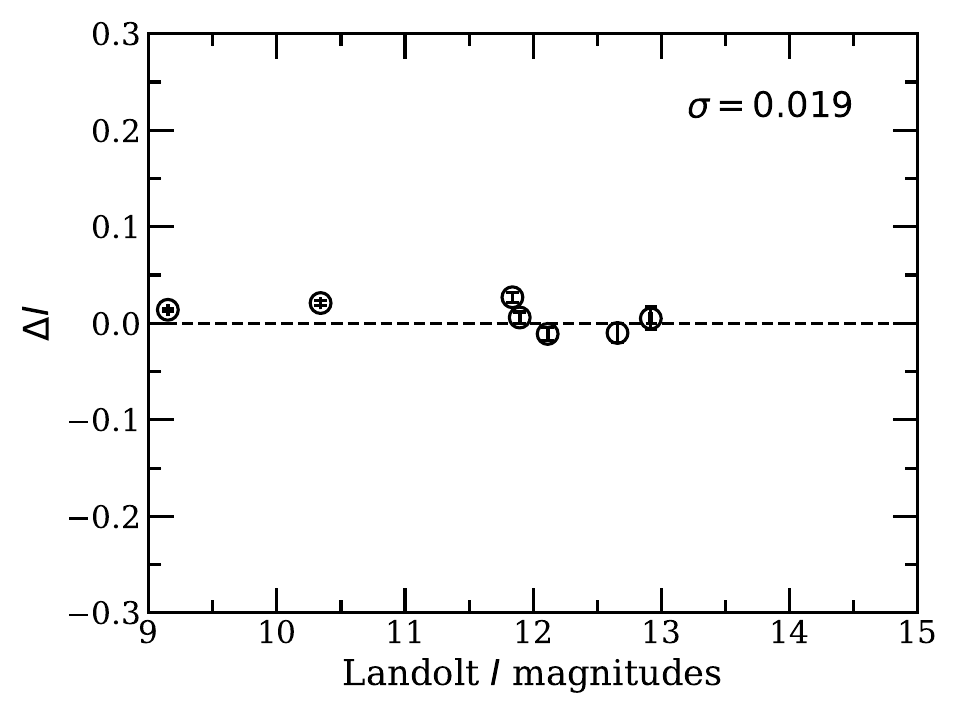} \\
                \includegraphics[width=0.32\textwidth]{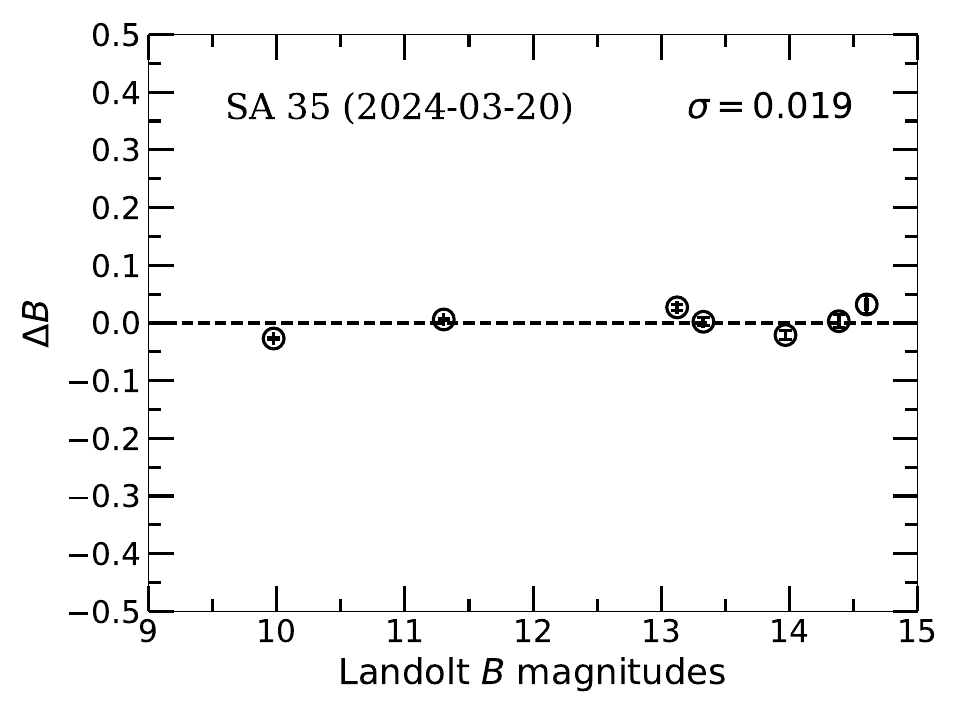} & \includegraphics[width=0.32\textwidth]{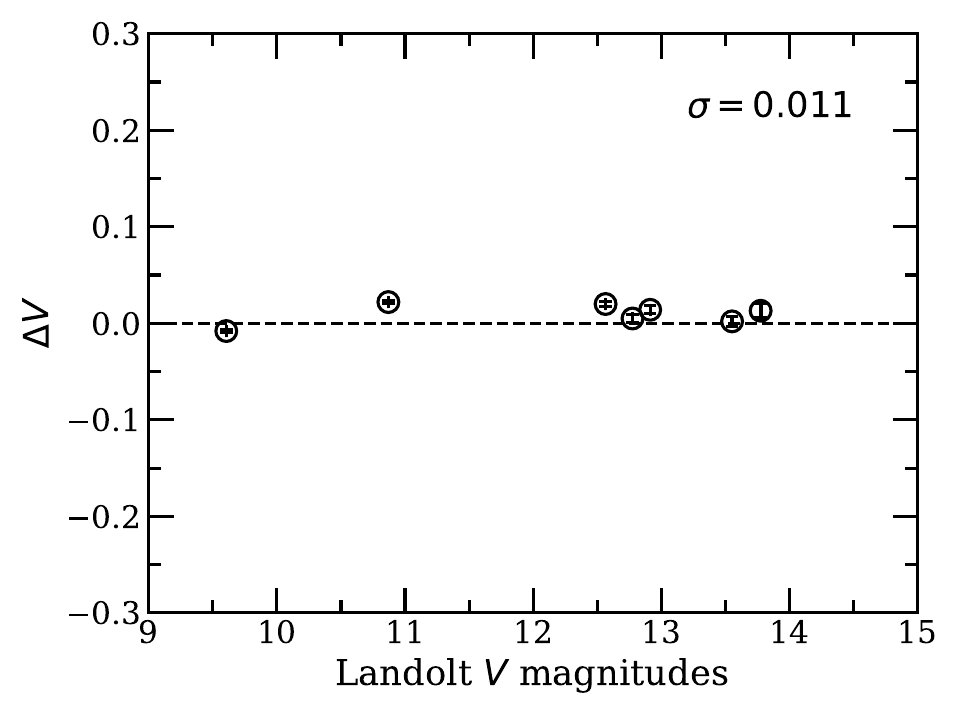} \\
                \includegraphics[width=0.32\textwidth]{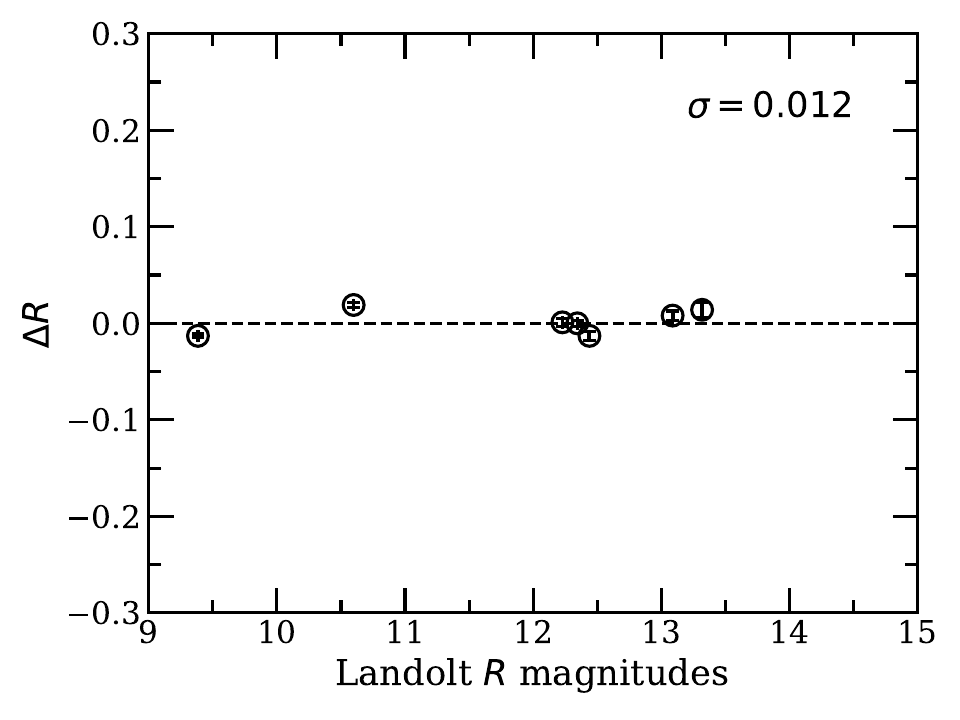} & \includegraphics[width=0.32\textwidth]{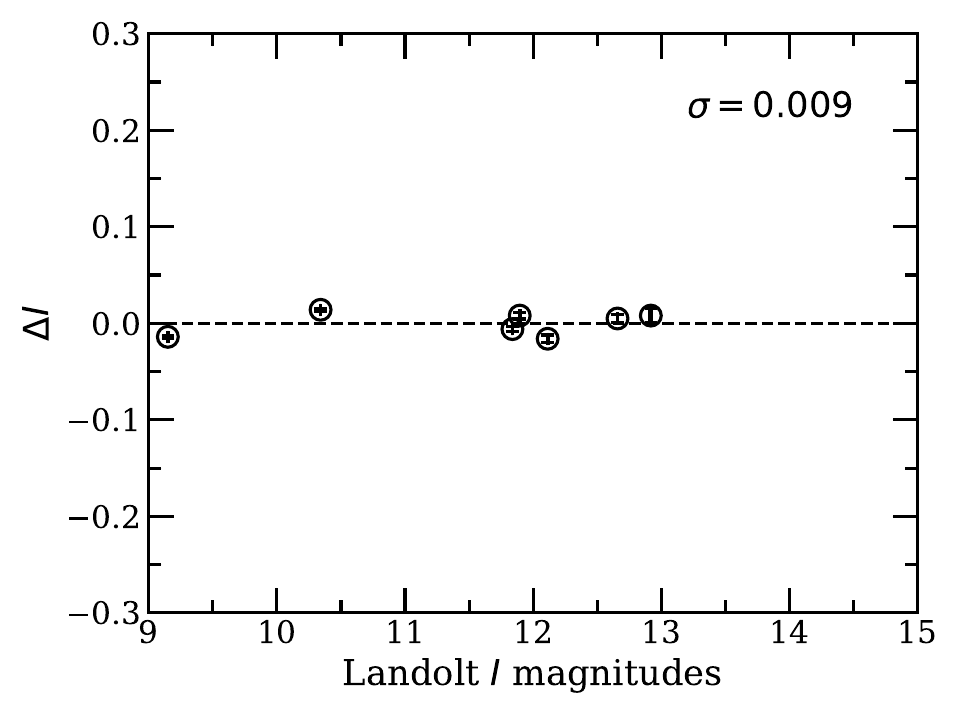} 
            \end{tabular}
            \caption{The residual plot of the transformation into the standard system.}
            \label{fig9}
        \end{figure*}

        Figure~\ref{fig9} shows the residuals of the measured magnitudes in each observing date. The extinction coefficients measured on 2023 February, 20th and 2024 March 20th are comparable to those measured in other Korean observatories (See Table~\ref{table4}). Considering the analysis in \citet{2008JASS...25..101K}, the heavy extinction observed on 2023 February 28th might be related to a considerable amount of fine dust in spring season or the possibility of thin cirrus clouds.

        \begin{deluxetable*}{ccccccc}[ht!]
                \tablecaption{The First-order atmospheric extinction coefficients in multiple filters for Korean observatories in units of mag per airmass.}
                \label{table4}
                \tablewidth{0pt}
                \tablehead{
                    \colhead{Site} &
                    \colhead{No Filter} &
                    \colhead{$B$} &
                    \colhead{$V$} &
                    \colhead{$R$} &
                    \colhead{$I$} &
                    \colhead{References}
                }
                \startdata
                CBNUO\tablenotemark{a} & $0.34$–$0.45$ & & & & & \citet{2008JASS...25..101K} \\
                BOAO\tablenotemark{b,c} & & $\sim0.30$ & $\sim0.20$ & $\sim0.10$ & $<0.10$ & \citet{1997PKAS...12..167K} \\
                SNU\tablenotemark{d} & & $0.36$–$0.90$ & $0.73$ & & & \citet{Lee+09}\\
                GAO\tablenotemark{e} & & $0.20$ & $0.13$ & & & \citet{2007JASS...24..209L}\\
                DOAO\tablenotemark{f} & & $0.88$ & $0.64$ & & & Kwon (2014)\tablenotemark{g}\\
                MAAO & & $0.48$–$1.38$ & $0.31$–$1.07$ & $0.22$–$0.75$ & $0.12$–$0.64$ & This work \\
                \enddata
                \tablecomments{}
                \tablenotetext{a}{Chungbuk National University Observatory}
                \tablenotetext{b}{Bohyunsan Optical Astronomy Observatory}
                \tablenotetext{c}{We adopted values at the nearest wavelength from Table 3 in \citet{1997PKAS...12..167K}.}
                \tablenotetext{d}{Department of Earth Science Education, Seoul National University.}
                \tablenotetext{e}{Gimhae Astronomical Observatory}
                \tablenotetext{f}{Deokheung Optical Astronomy Observatory}
                \tablenotetext{g}{Research Report 2014 of the National Youth Space Center (NYSC), \url{https://nysc.kywa.or.kr/organ/data.jsp}}
        \end{deluxetable*}
        
    \subsection{Limiting Magnitudes} \label{sec4.3}
        We measure limiting magnitudes of images in $BVRI$-band. A spiral galaxy NGC 3147 was observed in 10\,minutes exposure under a clear dark night on 2023 February 20th as shown in the left panel in Figure~\ref{fig10}. We perform the photometry using the \textsc{SExtractor} with a $2\times$FWHM aperture diameter. The FWHM values exhibit variation ranging from 2\farcs8 to 3\farcs8 with the filter and the focus. The photometric catalog from the Data Release 1 of Pan-STARRS\footnote{\url{https://catalogs.mast.stsci.edu/panstarrs/}} \citep[PS1;][]{2012ApJ...750...99T} is used as a reference catalog to calculate the zero-point magnitudes. We convert PS1 $gri$ magnitudes to $BVRI$ magnitudes (Refer Section 2.1 in \citealt{2023ApJ...949...33L}). As a result, the 5$\sigma$ limiting magnitude in $R$-band for the 10\,minutes exposure can be achieved as $\sim20$\,mag, which is comparable to that of other telescopes measured in Korea (Refer Table 1 in \citealt{2021JKAS...54...89I}). The other measured limiting magnitudes are summarized in Table~\ref{table5}.

        The $U$-band limiting magnitude is $16.8$\,AB mag. Due to the low sensitivity of the $U$-band, all the $U$-band images of SA 32 field obtained on 2023 Mar 29th with the total integrate time of 150\,minutes are stacked. The zero-point magnitude is measured using the magnitudes of standard stars from \citet{2013AJ....146..131L} and is converted into the AB system using \citet{2007AJ....133..734B}.

        After cleaning the mirror on 2023 May 30th, we additionally measured the limiting magnitudes of 5\,minutes exposure images of M101 field on 2023 June 16th. We calibrated the observed magnitudes with the same manner using the stars in the data release 9 (DR9) of the AAVSO photometric all-sky survey \citep{2016yCat.2336....0H} as comparison stars. The zero-points showed an improvement of $\sim0.4$ mag. 

        \begin{table*}[ht!]
            \centering
            \caption{The limiting magnitudes of images obtained from the MAAO 0.7\,m telescope in $UBVRI$-bands before and after cleaning the mirror.}
            \label{table5}
            \begin{tabular}{cccccccccc}
            \hline\hline
            Filter & Date-Obs (UT) & Zero Point (AB) & FWHM (\arcsec) & Exposures & 5$\sigma$ Limit (AB) \\
            \hline
                $U$ & 2023-03-29 13:01:26 & $16.63\pm0.17$ & $3.10\pm0.04$ & $600$\,s$\times5$+$300$\,s$\times20$ & $16.76$ \\
            \hline
                $B$ & 2023-02-20 14:55:26 & $20.87\pm0.07$ & $3.31\pm0.06$ & $120$\,s$\times5$ & $19.48$ \\
                $V$ & 2023-02-20 15:07:11 & $21.28\pm0.03$ & $3.84\pm0.06$ & & $19.36$ \\
                $R$ & 2023-02-20 15:18:53 & $21.44\pm0.02$ & $3.66\pm0.07$ & & $19.64$ \\
                $I$ & 2023-02-20 14:31:11 & $20.89\pm0.02$ & $2.85\pm0.06$ & & $19.59$ \\
            \hline
                $B$ & 2023-06-16 13:39:38 & $21.25\pm0.05$ & $2.59\pm0.04$ & $60$\,s$\times5$ & $18.08$ \\
                $V$ & 2023-06-16 13:45:41 & $21.66\pm0.04$ & $2.52\pm0.04$ & & $18.34$ \\
                $R$ & 2023-06-16 13:52:01 & $21.87\pm0.05$ & $2.58\pm0.08$ & & $18.53$ \\
                $I$ & 2023-06-16 13:58:04 & $21.16\pm0.03$ & $2.54\pm0.05$ & & $18.29$ \\   
            \hline
            \end{tabular}
        \end{table*}

\section{Current Scientific Programs} 
    We describe current research topics mainly focusing on time-series observational studies, including supernovae (SNe) and transiting exoplanets. 
    
\begin{figure*}[ht!]
      \centering
      \includegraphics[width=0.45\textwidth]{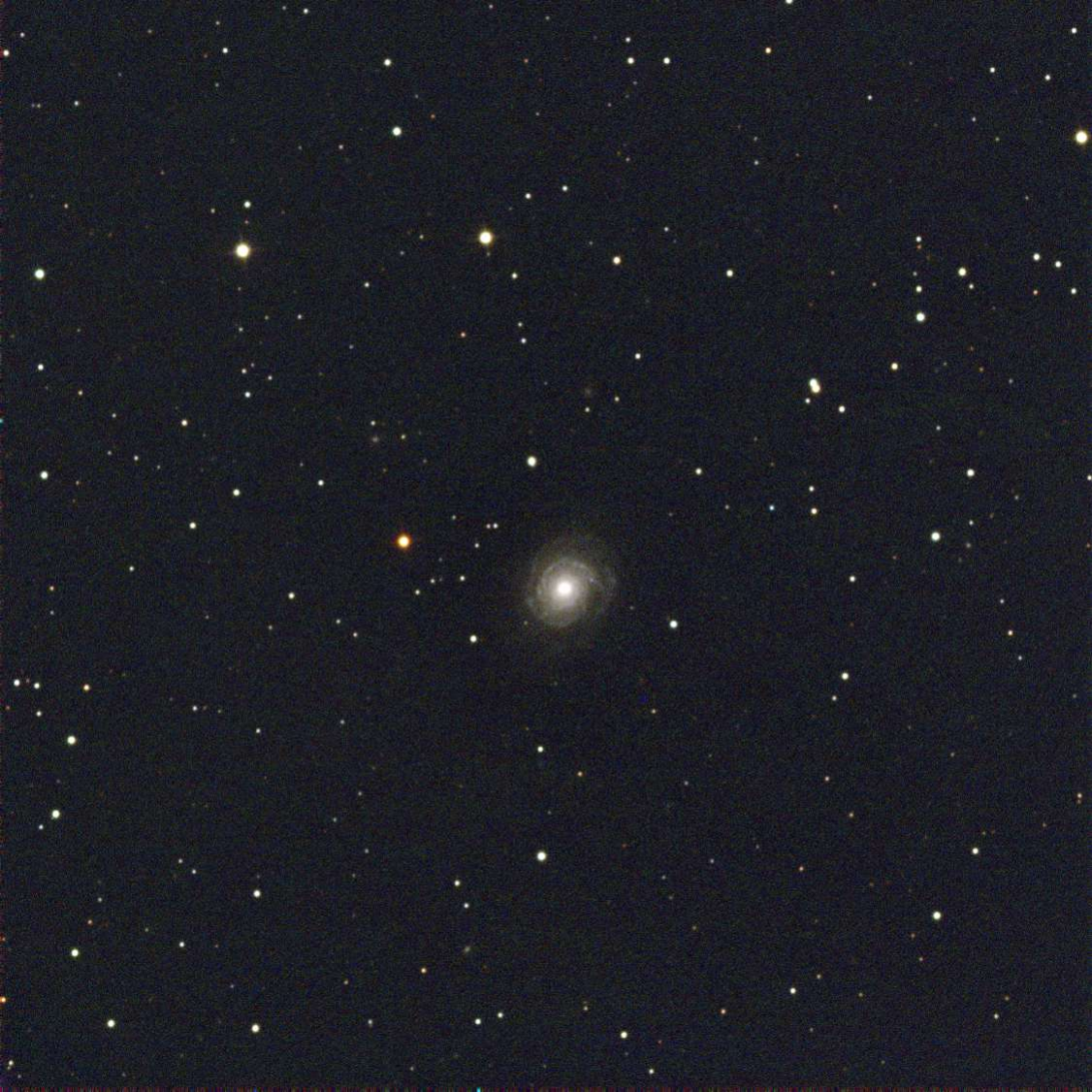}
      \hspace{0.01\textwidth}
      \includegraphics[width=0.45\textwidth]{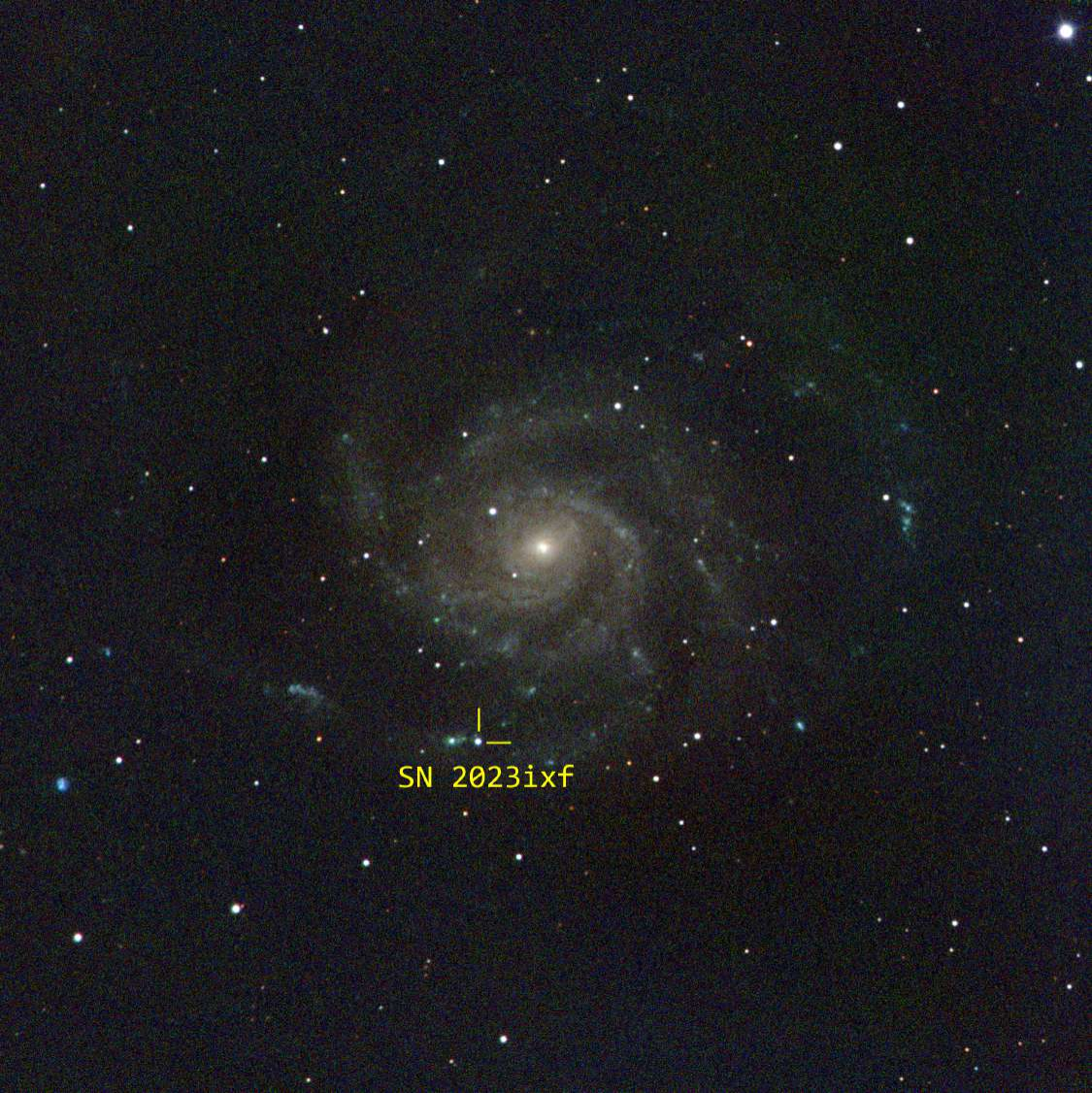}
      \caption{Sample images obtained from the 0.7\,m telescope. Left: $BVR$-bands color image for the ambiance of NGC 3147 used for measuring the limiting magnitudes on 2023 February 20th in Section~\ref{sec4.3}. Right: $VRI$-bands color image of M101 on 2023 May 20th. SN 2023ixf is marked in a yellow reticle.}
      \label{fig10}
    \end{figure*}

    (i) \textit{Supernovae}: The MAAO 0.7\,m telescope participates in a high-cadence monitoring program of nearby galaxies to obtain light curves of supernovae (SNe) in their infant phase, as part of a telescope network \citep{2019JKAS...52...11I, 2021JKAS...54...89I}. Monitoring the rising part of their light curves allows us to constrain SN progenitor models \citep{2015ApJS..221...22I,2023ApJ...949...33L}. One recent detection is SN 2023ixf (Type II), the nearest supernova in the last 10 yr. As soon as we received the transient alert, we captured the very early light of SN 2023ixf within a day of its discovery on 2023 May 19th \citep{2023TNSTR1158....1I} (see the right panel of Figure~\ref{fig10} for a sample $VRI$-band color image of M101, the host galaxy). This early data will contribute to our understanding of the fate of massive stars (Kim et al., in preparation).

    (ii) \textit{Transiting Exoplanets}: Space-based telescopes searching for transiting exoplanets such as Kepler \citep{2010Sci...327..977B} and TESS \citep{2015JATIS...1a4003R} have monitored a large number of stars with high-cadence and high photometric precision for many years, discovering more than thousands of the exoplanets and objects of interests showing planetary features. The light curves and stellar and planetary parameters of these objects are published in databases (e.g., NASA Exoplanet Archive; \citealt{2013PASP..125..989A}). However, due to the limitation of their lifetime and originally designed mission, the targets have been monitored with a limited time duration and filters. So their photometric follow-up observation with small telescopes can allow us to confirm them as exoplanets and characterize their physical properties (mass, radius, density, etc.) and orbital ephemerides. We are now performing a pilot observation of a confirmed transiting exoplanet to examine the capabilities of MAAO.
    \label{sec5}

\section{Future work} \label{sec6}
    Future work involves a long-term monitoring of observing conditions and system stabilization. The study of the seasonal change of the astronomical seeing (Lim et al., in preparation), atmospheric extinction coefficients, and the influence of light pollution on the sky brightness at MAAO (Park et al., in preparation) are ongoing. These studies would provide observers with insights into establishing observation strategies, and reference materials for MAAO staff to maintain the instrument and educate the public in consideration of local weather conditions.    

\section{Summary and Conclusion} \label{sec7}
     We present the performance of the 0.7\,m robotic telescope system at MAAO and its standard photometric system, and current science programs. The 0.7\,m telescope is newly installed at MAAO, a public observatory in Miryang, Korea. We have established a robotic observing system to utilize this system for scientific observations following public events. The gain, readout noise, residual signal, dark current of the detector, and the total throughput are also evaluated. We find that the PSF shape is spatially uniform across the overall field of view with an ellipticity of $<$0.1 for a 1\,minutes exposure time. Subsequently, photometric calibration is performed using standard stars in the $BVRI$-band. The atmospheric extinction coefficients are moderate compared to those of other observatories in Korea, but we find severe extinction which needs to be confirmed with future data. The 0.7\,m telescope can achieve image depths down to $BVRI$$\sim$19.4–19.6\,AB mag at $5\sigma$ with a 10\,minutes integrated time under clear dark sky conditions and a seeing condition of $\sim$$2\farcs9$-$3\farcs8$, comparable to those of other Korean observatories used for research. Due to the low system sensitivity in ultraviolet, the $U$-band depth is shallow ($U$$\sim$$16.8$\,AB mag) even with a 150\,minutes exposure time. While the MAAO system serves as a public facility, our automation efforts make it available for astronomical research, especially for time-domain astronomy such as, supernova monitoring and transiting exoplanets. In the future, we plan to stabilize the automated observation.
\begin{acknowledgments}
    This work acknowledges support from the Basic Science Research Program through the National Research Foundation of Korea (NRF) funded by MSIT (No. 2022R1A6A3A01085930). D.K. acknowledges support by the NRF of Korea (NRF) grant (No. 2021R1C1C1013580 and 2022R1A4A3031306) funded by MSIT. M.I. and H.C. acknowledge the support from the National Research Foundation of Korea (NRF) grants, No. 2020R1A2C3011091, and No. 2021M3F7A1084525, funded by the Korea government (MSIT). We sincerely thank the MAAO staff and Metaspace, Inc. for the kind assistance and maintenance of the facilities. The authors also thank Bill Dean of PlaneWave Instruments for providing us with the response curve of the telescope.
\end{acknowledgments}

%
\vspace{5mm}
\facilities{MAAO:0.7m}
\software{Astrometry.net \citep{2010AJ....139.1782L}, cloudynight \citep{2020AJ....159..178M}, SExtractor \citep{1996A&AS..117..393B}, PSFEx \citep{2013ascl.soft01001B}, PyRAF \citep{2012ascl.soft07011S}, WeatherLink APIv2 (\url{https://weatherlink.github.io/v2-api/}), pywin32 (\url{https://github.com/mhammond/pywin32})}

{}



\end{document}